%% file: main.tex
\PassOptionsToPackage{dvipsnames, table}{xcolor}
\documentclass[conference]{IEEEtran}
\IEEEoverridecommandlockouts

\usepackage{cite}
\usepackage{amsmath,amssymb,amsfonts}
\usepackage{algorithmic}
\usepackage{graphicx}
\usepackage{textcomp}
\def\BibTeX{{\rm B\kern-.05em{\sc i\kern-.025em b}\kern-.08em
    T\kern-.1667em\lower.7ex\hbox{E}\kern-.125emX}}

\usepackage{color,xcolor}
\usepackage[table]{xcolor}
\usepackage{booktabs}
\usepackage{xspace}
\usepackage{newtxmath}
\usepackage{balance}
\usepackage{rotating}
\usepackage{multirow}
\usepackage{url}
\usepackage[colorlinks,citecolor=citecolor, linkcolor=linkcolor, bookmarks=false]{hyperref}
\definecolor{citecolor}{HTML}{1F801F}
\definecolor{linkcolor}{HTML}{ED1C24}

\newcommand{\method}{HCMEN\xspace}

\definecolor{maroon}{HTML}{800000}

\begin{document}

\title{Hybrid CNN-Mamba Enhancement Network for Robust Multimodal Sentiment Analysis}

\author{
Xiang Li$^1$, Xianfu Cheng$^2$, Xiaoming Zhang$^{1,\dagger}$\thanks{$\dagger$: Corresponding author.}, Zhoujun Li$^2$\\
$^1$ School of Cyber Science and Technology, Beihang University, Beijing\\
$^2$School of Computer Science and Engineering, Beihang University, Beijing \\
\{xlggg, buaacxf, yolixs, lizj\}@buaa.edu.cn
}

\maketitle

\input{Section/0_abs}

\input{Section/1_intro}
\input{Section/2_method}

\input{Section/3_exp}
\input{Section/4_summary}

% \clearpage
% \newpage
\bibliographystyle{IEEEbib}  % 可根据需要改为 IEEEtran、plain、unsrt 等
\bibliography{references}    % 对应的是 references.bib 文件名
\end{document}

%% file: Section/0_abs.tex
\begin{abstract}
Multimodal Sentiment Analysis (MSA) with missing modalities has recently attracted increasing attention. Although existing research mainly focuses on designing complex model architectures to handle incomplete data, it still faces significant challenges in effectively aligning and fusing multimodal information.
In this paper, we propose a novel framework called the Hybrid CNN-Mamba Enhancement Network (HCMEN) for robust multimodal sentiment analysis under missing modality conditions. HCMEN is designed around three key components: (1) hierarchical unimodal modeling, (2) cross-modal enhancement and alignment, and (3) multimodal mix-up fusion.
First, HCMEN integrates the strengths of Convolutional Neural Network (CNN) for capturing local details and the Mamba architecture for modeling global contextual dependencies across different modalities. Furthermore, grounded in the principle of Mutual Information Maximization, we introduce a cross-modal enhancement mechanism that generates proxy modalities from mixed token-level representations and learns fine-grained token-level correspondences between modalities.
The enhanced unimodal features are then fused and passed through the CNN-Mamba backbone, enabling local-to-global cross-modal interaction and comprehensive multimodal integration.
Extensive experiments on two benchmark MSA datasets demonstrate that HCMEN consistently outperforms existing state-of-the-art methods, achieving superior performance across various missing modality scenarios.
The code will be released publicly in the near future.
\end{abstract}

%% file: Section/1_intro.tex
\section{Introduction}
Multimodal Sentiment Analysis (MSA) aims to infer a speaker's emotional state by integrating diverse modalities such as language, vision, and audio \cite{yang2022disentangled,mai2022hybrid,zhang2023learning,li2025learning}. This task plays a crucial role in human-centered applications, including human-computer interaction, opinion mining, and mental health analysis. Despite its potential, MSA remains challenging due to two core issues: (1) the high variability and noise across modalities in real-world data, and (2) the difficulty of modeling complex cross-modal dependencies, particularly under missing or incomplete modality conditions.

Recent studies have leveraged multimodal fusion strategies to improve the robustness and accuracy of sentiment prediction. Existing methods can be broadly categorized into three types: feature-level, decision-level, and model-level fusion. Feature-level fusion concatenates modality-specific features to form a joint representation~\cite{williams2018recognizing,cai2020feature}, while decision-level fusion aggregates independent predictions from each modality~\cite{wu2010emotion,zhang2023adamow}. More recently, model-level fusion has emerged as a powerful approach to learn dynamic inter- and intra-modal relationships~\cite{tsai2019multimodal,yuan2021transformer,sun2022cubemlp,sun2023efficient,wang2023tetfn,li2023decoupled,li2024toward,li2024adaptive,zeng2024disentanglement}, often employing Transformer-based architectures to capture long-range dependencies. While Transformer-based models offer strong expressive power, they often suffer from high computational complexity and limited efficiency when processing long sequences.

Recently, Mamba \cite{gu2022parameterization,gu2023mamba}, a selective state space model, has emerged as a promising alternative due to its linear-time sequence modeling and superior efficiency in capturing long-range dependencies.
Studies such as~\cite{li2024coupled, ye2025depmamba, li2025tf} have demonstrated the potential and applicability of the Mamba architecture in multimodal fusion and sentiment analysis tasks.
Despite these advantages, existing Mamba-based fusion methods focus primarily on global modeling and overlook the need for fine-grained cross-modal alignment. This oversight hampers their scalability and robustness, especially in real-world scenarios where modalities are noisy or partially missing.
\begin{figure*}[!t]
    \centering
    \includegraphics[width=1\linewidth]{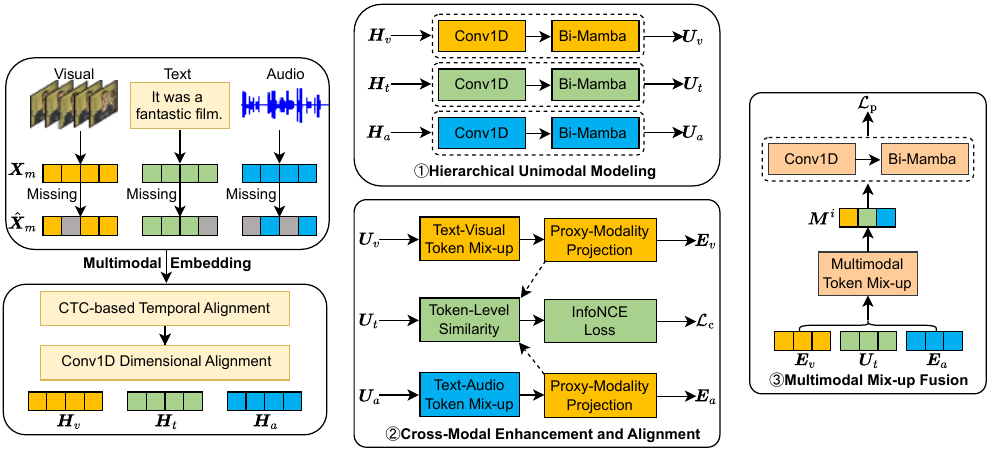}
    \caption{Architecture of the Hybrid CNN-Mamba Enhancement Network.}
    \label{fig:main}
\end{figure*}
To address these challenges, we propose a novel architecture termed \textbf{\method}, a Hybrid CNN-Mamba Enhancement Network designed for robust and efficient multimodal sentiment analysis. Our method introduces three key innovations:
(1) \textit{Hierarchical Contextual Modeling} serves as a hybrid unimodal backbone that integrates CNNs for local pattern extraction and Mamba-based state space models for efficient long-range unimodal dependency modeling.
(2) \textit{Cross-modal Enhancement and Alignment} constructs proxy representations across modalities (e.g., audio-to-text, visual-to-text) based on mutual information maximization and structured contrastive learning, and aligns them at the token level via averaged cosine similarity.
(3) \textit{Multimodal Mix-up Fusion} interleaves aligned vision, audio, and text tokens to create hybrid token sequences that simulate diverse modality combinations, which are then fed into the CNN-Mamba backbone for progressive token-level multimodal modeling.
Extensive experiments on benchmark sentiment datasets (e.g., CMU-MOSI and CMU-MOSEI) demonstrate that \method achieves state-of-the-art performance while being more parameter- and computation-efficient compared to Transformer-based methods.

\noindent\textbf{Contributions.} This work introduces \method, the first hybrid CNN-Mamba architecture for robust multimodal sentiment analysis, effectively handling missing or corrupted modalities. By combining CNN for local feature extraction with Mamba for efficient long-range modeling, \method enables progressive cross-modal enhancement through structured contrastive alignment and token-level mix-up fusion. Extensive experiments on benchmark datasets confirm its superiority over existing methods in both performance and efficiency.

%% file: Section/2_method.tex
\section{Proposed Method}
\subsection{Preliminary of Mamba}
In recent years, {State Space Models (SSMs)} have witnessed significant advancements~\cite{gu2022parameterization,gu2023mamba}. Originating from classical control theory, SSMs offer an effective framework for modeling long-range dependencies with linear computational complexity. These models introduce a hidden state $\boldsymbol{h}(t)\in\mathbb{R}^{N}$ to transform the input $\boldsymbol{x}(t)\in\mathbb{R}^{L}$ into the output $\boldsymbol{y}(t)\in\mathbb{R}^{L}$, where $N$ and $L$ represent the number of hidden states and the sequence length, respectively. The continuous-time dynamics of an SSM can be described by:
\begin{equation}
\label{eq1}
\begin{aligned}
    \boldsymbol{h'}(t) = \mathbf{A}\boldsymbol{h}(t) + \mathbf{B}\boldsymbol{x}(t), \,\,\,
    \boldsymbol{y}(t) = \mathbf{C}\boldsymbol{h}(t). 
\end{aligned}
\end{equation}
Here, $\mathbf{A} \in \mathbb{R}^{N \times N}$ is the state transition matrix, and $\mathbf{B}, \mathbf{C} \in \mathbb{R}^{N \times 1}$ denote the input and output projection matrices, respectively. The {Mamba} architecture further extends this framework by introducing a time-step parameter $\Delta$, enabling discretization of the continuous parameters $\mathbf{A}$ and $\mathbf{B}$ into $\mathbf{\overline{A}}$ and $\mathbf{\overline{B}}$ using the \textit{zero-order hold (ZOH)} method. Specifically, the discretized matrices are given by $\mathbf{\overline{A}} = \exp(\mathbf{\Delta A})$ and $\mathbf{\overline{B}} = (\mathbf{\Delta A})^{-1} (\exp(\mathbf{\Delta A}) - \mathbf{I}) \cdot \mathbf{\Delta B}$. With this, the continuous system in Eq.~\eqref{eq1} can be reformulated in discrete-time recurrent form:
\begin{equation}
\label{eq3}
\begin{aligned}
    \boldsymbol{h}_{t} = \mathbf{\overline{A}}\boldsymbol{h}_{t-1}+\mathbf{\overline{B}}\boldsymbol{x}_{t}, \,\,\,
    \boldsymbol{y}_{t} = \mathbf{C}\boldsymbol{h}_{t}.
\end{aligned}
\end{equation}
Moreover, Eq.~\eqref{eq3} can be equivalently expressed in convolutional form:
\[
\mathbf{\overline{K}} = \left( \mathbf{C\overline{B},\ldots,C\overline{A}^{\mathrm{L}-1}\overline{B}} \right), \quad \boldsymbol{y} = \boldsymbol{x}\circledast\mathbf{\overline{K}},
\]
where $\circledast$ denotes the convolution operation and $\mathbf{\overline{K}} \in \mathbb{R}^{L}$ is the global convolution kernel. {Mamba} significantly improves deep sequence modeling by leveraging its data-dependent and computationally efficient design.
In this paper, we adopt the {bi-directional Mamba (Bi-Mamba)}~\cite{liu2024vmamba} as our baseline model, which captures contextual information from both forward and backward directions, thereby enabling a more comprehensive understanding of long-range dependencies.
\subsection{Overview of \method}
As illustrated in Fig.~\ref{fig:main}, we propose the Hybrid CNN-Mamba Enhancement Network (HCMEN) for robust multimodal sentiment analysis under incomplete modality conditions. HCMEN starts by projecting pre-extracted unimodal features from public datasets into a shared latent space, forming unified multimodal embeddings. The framework comprises three key components that facilitate effective cross-modal alignment and interaction:
(i) Hierarchical Unimodal Modeling applies 1D convolutional layers to extract local semantics, followed by Mamba modules for efficient long-range dependency modeling, enabling hierarchical representation from local to global levels.
(ii) Cross-modal Enhancement and Alignment leverages mutual information maximization to generate proxy representations from mixed token-level inputs and learn structured correspondences across modalities, promoting semantic alignment and information sharing.
(iii) Multimodal Mix-up Fusion feeds enhanced unimodal features into a CNN-Mamba backbone, combining CNNs for local modeling with Mamba for global reasoning. A progressive fusion strategy with stacked hybrid blocks enables deep cross-modal interaction across layers.
The final fused representation is passed to a linear classification head for sentiment prediction.
\subsection{Multimodal Input Embedding}
Following prior work~\cite{yuan2021transformer,zhang2024towards}, we use pre-extracted features for each modality from benchmark datasets. For modality \( m \in \{t, v, a\} \) (text, visual, acoustic), the input is denoted as \(\mathbf{X}_m \in \mathbb{R}^{T_m \times D_m}\), where \(T_m\) is the sequence length and \(D_m\) the feature dimension. To simulate incomplete modality scenarios, we apply random masking or substitution to obtain corrupted inputs \(\widehat{\mathbf{X}}_m\), following the approach in LNLN~\cite{zhang2024towards}. This enables the model to handle varying modality availability.

Each corrupted sequence is aligned using a CTC-based block \cite{tsai2019multimodal}, then projected via a 1D convolution: 
$
\{\mathbf{H}_t, \mathbf{H}_v, \mathbf{H}_a\} = \mathrm{Conv1D}(\mathrm{CTC}(\{\widehat{\mathbf{X}}_t, \widehat{\mathbf{X}}_v, \widehat{\mathbf{X}}_a\})).
$
It maps all modalities into a latent space of fixed length \(L\) and dimension \(D\), producing unified embeddings \(\mathbf{H}_m \in \mathbb{R}^{L \times D}\) for subsequent fusion.
\subsection{Hierarchical Unimodal Modeling}
\label{sec:unimodal_modeling}
To capture both short- and long-range temporal patterns within each modality, we adopt a hierarchical modeling strategy. Given the aligned input \(\mathbf{H}_{m}\) for modality \(m \in \{t, v, a\}\), we first extract local representations using Layer Normalization (LN), a depth-wise 1D convolution, and a residual connection:
\begin{equation}
    \mathbf{H}^{{local}}_{m} = \mathbf{H}_{m} + \mathrm{Conv1D}(\mathrm{LN}(\mathbf{H}_{m})).
\end{equation}
To model global dependencies, we apply another LN and a Bi-Mamba module, again with residual connection:
\begin{equation}
    \mathbf{H}^{{global}}_{m} =  \mathbf{H}^{{local}}_{m} + \text{Bi-Mamba}(\mathrm{LN}( \mathbf{H}^{{local}}_{m})).
\end{equation}
The resulting embedding is defined as $\mathbf{U}_{m} = \mathbf{H}^{{global}}_{m}$.
This hierarchical module combines fine-grained and contextual cues, enabling each modality to generate expressive representations for subsequent multimodal enhancement.
\subsection{Cross-Modal Enhancement and Alignment}
\label{sec:cross_modal_alignment}

To tackle missing or corrupted modalities, we introduce the {Cross-Modal Enhancement and Alignment (CMEA)} module. CMEA generates proxy representations for weaker modalities and aligns them with the text modality by maximizing token-level semantic consistency.

Given unimodal features \(\mathbf{U}_{m} \in \mathbb{R}^{L \times D}\) for \(m \in \{t, v, a\}\), we enhance the visual and acoustic modalities by probabilistically mixing them with corresponding text tokens:
\begin{equation}
    \widehat{\mathbf{U}}_{m}^i = 
    \begin{cases}
        \mathbf{U}_{t}^i, & \text{with probability } p > p*, \\
        \mathbf{U}_{m}^i, & \text{otherwise}, \quad m \in \{v, a\}.
    \end{cases}
\end{equation}

The mixed features \(\widehat{\mathbf{U}}_{m}\) are transformed into proxy representations via a modality-specific MLP:
\begin{equation}
    {\mathbf{E}}_{m} = \text{MLP}_m( \widehat{\mathbf{U}}_{m}), \quad m \in \{v, a\}.
\end{equation}

To align each \({\mathbf{E}}^{m}\) with the corresponding text embedding \(\mathbf{U}^{t}\), we compute the average token-wise cosine similarity:
\begin{equation}
    \text{sim}({\mathbf{E}}_{m}, \mathbf{U}_t) = \frac{1}{L} \sum_{i=1}^{L} \frac{ {\mathbf{E}}_m^i \cdot \mathbf{U}_t^i }{ \| {\mathbf{E}}_m^i \|_2 \| \mathbf{U}_t^i \|_2 }.
\end{equation}

An InfoNCE loss is then applied to maximize mutual
information  with the matched text while suppressing similarities with other negatives:
\begin{equation}
    \mathcal{L}_{\text{c}} = \frac{1}{2}  \sum_{m \in \{v, a\}} (\frac{1}{B} \sum_{i=1}^{B} 
    -\log \frac{
        \exp \left( \text{sim}({\mathbf{E}}_{m;i}, \mathbf{U}_{t;i})/\tau \right)
    }{
        \sum_{j=1}^{B} \exp \left(\text{sim}({\mathbf{E}}_{m;i}, \mathbf{U}_{t;j})/\tau \right)
    }),
\end{equation}
where \(B\) is the batch size and \(\tau\) is the temperature.

By injecting semantic priors from text and enforcing alignment at the token level, CMEA enhances robustness to modality degradation and promotes multimodal fusion.

\subsection{Multimodal Mix-up Fusion}
\label{sec:mm_mixup_fusion}
After unimodal enhancement and cross-modal alignment, we introduce a {Multimodal Mix-up Fusion} module to perform deep integration across modalities.

Given the aligned features $\mathbf{U}_{t}, {\mathbf{E}}_{v}, {\mathbf{E}}_{a} \in \mathbb{R}^{L \times D}$, we construct the fused sequence $\mathbf{M} \in \mathbb{R}^{3L \times D}$ by interleaving tokens from each modality at every time step:
\begin{equation}
    \mathbf{M} = [ \mathbf{E}^1_v, \mathbf{U}^1_t, \mathbf{E}^1_a, \dots, \mathbf{E}^L_v, \mathbf{U}^L_t, \mathbf{E}^L_a ].
\end{equation}
This interleaving preserves fine-grained temporal alignment and enables tightly coupled cross-modal interactions.

As in unimodal modeling, we adopt a progressive fusion backbone composed of stacked hybrid blocks to perform deep multimodal integration. Each block consists of two sequential stages: a local CNN modeling stage, where a LayerNorm followed by a 1D convolution captures short-range dependencies, and a global Mamba reasoning stage, where another LayerNorm and a Mamba layer efficiently model long-range sequential interactions. Specifically, each fusion block updates the representation as:
\begin{align}
\mathbf{F}_z^{local} &= \mathbf{M} + \text{Conv1D} \left( \text{LN}(\mathbf{M}) \right) , \\
\mathbf{F}_z^{global} &= \mathbf{F}_z^{local} + \text{Bi-Mamba} ( \text{LN}({\mathbf{F}_z^{local}}) ).
\end{align}
This design seamlessly combines CNN’s strength in capturing localized features with Mamba’s ability to reason over global context, enabling expressive and efficient multimodal fusion.

\input{Table/SOTA}
\subsection{Training and Optimization}
\label{sec:training_optimization}
To derive the utterance-level representation, we apply mean pooling over the fused sequence $\mathbf{F}_z^{{global}}$:
\begin{equation}
    \mathbf{F}^{h} = \text{Mean}(\mathbf{F}_z^{{global}}) \in \mathbb{R}^{D}
\end{equation}
A fully connected layer is then used to predict the final sentiment score: $\hat{y} = \text{FC}(\mathbf{F}^{h})$. The primary objective is a sentiment prediction loss, measured by Mean Squared Error (MSE) between the predicted and ground-truth values:
    \begin{equation}
        \mathcal{L}_{{p}} = \| \hat{y} - y \|_2^2.
    \end{equation}

To encourage cross-modal consistency, we further introduce a token-level contrastive alignment loss $\mathcal{L}_{c}$. The overall training objective combines both terms:
\begin{equation}
\mathcal{L}_{\text{total}} = \mathcal{L}_{p} + \alpha \cdot \mathcal{L}_{c},
\end{equation}
where $\alpha$ controls the trade-off between sentiment prediction and modality alignment.

%% file: Table/SOTA.tex
\begin{table*}[t]
\centering
\caption{Overall performance comparison on the MOSI and MOSEI datasets under missing modality settings.}
\label{tab:SOTA}
\resizebox{\textwidth}{!}{ 
\begin{tabular}{ccccccccccccc}
\toprule
\multirow{2}{*}{Method} & \multicolumn{6}{c}{MOSI} & \multicolumn{6}{c}{MOSEI} \\
\cmidrule(lr){2-7} \cmidrule(lr){8-13}
 & Acc-7 & Acc-5 & Acc-2 & F1 & MAE & Corr & Acc-7 & Acc-5 & Acc-2 & F1 & MAE & Corr \\
\midrule
MISA \cite{hazarika2020misa} & 29.85 & 33.08 & 71.49 / 70.33 & 71.28 / 70.00 & 1.085 & 0.524 & 40.84 & 39.39 & 71.27 / 75.82 & 63.85 / 68.73 & 0.780 & 0.503 \\
Self-MM \cite{yu2021learning} & 29.55 & 34.67 & 70.51 / 69.26 & 66.60 / 67.54 & 1.070 & 0.512 & 44.70 & 45.38 & 73.89 / 77.42 & 68.92 / 72.31 & 0.695 & 0.498 \\
MMIM \cite{han2021improving}& 31.30 & 33.77 & 69.14 / 67.06 & 66.65 / 64.04 & 1.077 & 0.507 & 40.75 & 41.74 & 73.32 / 75.89 & 68.72 / 70.32 & 0.739 & 0.489 \\
TFR-Net \cite{yuan2021transformer} & 29.54 & 34.67 & 68.15 / 66.35 & 61.73 / 60.06 & 1.200 & 0.459 & 46.83 & 34.67 & 73.62 / 77.23 & 68.80 / 71.99 & 0.697 & 0.489 \\
CENET \cite{wang2022cross} & 30.38 & 33.62 & 71.46 / 67.73 & 68.41 / 64.85 & 1.080 & 0.504 & \textbf{47.18} & \textbf{47.83} & 74.67 / 77.34 & 70.68 / 74.08 & 0.685 & 0.535 \\
ALMT \cite{zhang2023learning}& 30.30 & 33.42 & 70.40 / 68.39 & 72.57 / 71.80 & 1.083 & 0.498 & 40.92 & 41.64 & 76.64 / 77.54 & 77.14 / 78.03 & 0.674 & 0.481 \\
BI-Mamba \cite{yang2024cardiovascular}& 31.20 & 34.02 & 71.74 / 71.12 & 71.83 / 71.11 & 1.087 & 0.498 & 45.12 & 45.76 & 76.82 / 76.72 & 76.35 / 76.38 & 0.701 & 0.545 \\
LNLN \cite{zhang2024towards} & 32.53 & 36.25 & 71.91 / 70.11 & 71.71 / 70.02 & 1.062 & 0.503 & 45.42 & 46.17 & 76.30 / {78.19} & {77.77} / \textbf{79.95} & 0.692 & 0.530 \\
\midrule
\textbf{HCMEN} & \textbf{34.37} & \textbf{38.12} & \textbf{74.79} / \textbf{73.50} & \textbf{74.78} / \textbf{73.41} & \textbf{1.034} & \textbf{0.546} 
& 46.17 & \underline{46.92} & \textbf{78.14} / \textbf{78.30} & \textbf{78.11} / 76.93 & \textbf{0.662} & \textbf{0.599} \\
\bottomrule
\end{tabular}%
}

\end{table*}

%% file: Section/3_exp.tex
\section{Experiments}
\subsection{Experimental Setup}
\noindent\textbf{Datasets and Metrics.}\quad
We evaluate our method on two standard MSA benchmarks: {CMU-MOSI}\cite{zadeh2016multimodal} and {CMU-MOSEI}\cite{zadeh2018multimodal}, using the {unaligned} setting with publicly available pre-extracted features.
CMU-MOSI contains 2,199 English video segments labeled on a 7-point sentiment scale, split into 1,284 for training, 229 for validation, and 686 for testing. CMU-MOSEI includes 22,856 utterances from over 1,000 speakers, with standard splits of 16,326/1,871/4,659 for training, validation, and testing.
Following Zhang et al.~\cite{zhang2024towards}, we adopt both classification and regression metrics: Acc-7 and Acc-5 for multi-class accuracy, Acc-2 and F1 for binary sentiment classification, MAE for prediction error, and Pearson correlation (Corr) for prediction consistency. Higher values indicate better performance, except for MAE.

\subsection{Comparison Results}
As shown in Table~\ref{tab:SOTA}, our proposed \method{} consistently outperforms state-of-the-art methods across all metrics on both the MOSI and MOSEI datasets. Notably, it achieves the highest average F1 scores—74.78 on MOSI and 78.11 on MOSEI—demonstrating strong sentiment prediction capability. \method{} also excels in MAE and Corr, indicating its effectiveness in capturing subtle emotional cues with lower prediction error. Compared to Transformer-based models such as TFR-Net, ALMT, and LNLN, our method offers better efficiency and robustness, owing to its hierarchical CNN-Mamba architecture. Furthermore, the cross-modal enhancement and alignment modules significantly boost inter-modal fusion, enabling accurate and resilient sentiment inference even under missing modality conditions. These results highlight the strength of our hybrid modeling, progressive fusion, and cross-modal augmentation strategies in robust MSA.
\input{Table/Ablation}
\subsection{Ablation Study}
We perform ablation studies on the MOSI dataset to evaluate the impact of key components in our model: CNN (local temporal modeling), Mamba (global sequence modeling), and CEMA (cross-modal enhancement and alignment). As shown in Table~\ref{tab:ablation}, removing any component degrades performance.
Specifically, removing CNN slightly reduces F1 (-1.16) and Acc-7 (-0.55), suggesting that local patterns are helpful but less critical. In contrast, removing Mamba causes a larger drop (-1.94 F1, -1.46 Acc-7), highlighting the importance of long-range temporal modeling. Excluding CEMA leads to the worst MAE (1.080) and the largest F1 decrease (-2.17), underscoring its key role in robust cross-modal fusion.
These findings validate the effectiveness of our hybrid CNN-Mamba backbone with the crossmodal enhancement strategy.
\subsection{Efficiency Analysis}
Our model employs a hybrid CNN-Mamba backbone, offering a lightweight and scalable alternative to Transformer-based SOTA architectures. While Transformers suffer from quadratic complexity with respect to sequence, our backbone achieves linear growth, enabling more efficient long-range modeling. Under comparable configurations and excluding the influence of pre-trained encoders, our model reduces parameter overhead by approximately 60\% compared to the state-of-the-art Transformer baseline, highlighting its computational efficiency. These results confirm the advantage of our design in both performance and practical deployment for robust MSA.

%% file: Table/Ablation.tex
\begin{table}[t]
    \centering
    \caption{Ablation study. ‘w/o’ denotes removing the component.}
    \label{tab:ablation}
    \begin{tabular}{cccc}
        \toprule
        {MOdel}& {MAE} & {F1} & {ACC-7} \\
        \midrule
        w/o CNN & 1.061 & 73.62 & 33.82  \\
        w/o Mamba & 1.074 & 72.84 & 32.91  \\
         w/o CEMA & 1.080  & 72.61 & 33.92  \\
        \midrule
        \textbf{\method} &\textbf{1.034} & \textbf{74.78 }  & \textbf{34.37 }  \\
        \bottomrule
    \end{tabular}
\end{table}

%% file: Section/4_summary.tex
\section{Conclusion}
We propose HCMEN, the first hybrid CNN-Mamba framework for robust multimodal sentiment analysis under missing modality conditions. By integrating local feature extraction with efficient global modeling and introducing cross-modal enhancement based on mutual information maximization, HCMEN achieves effective alignment and fusion of incomplete modalities. Extensive experiments demonstrate its superior performance over state-of-the-art methods, validating its potential for practical applications.